\documentclass{vgtc}                          

\ifpdf
  \pdfoutput=1\relax                   
  \usepackage{graphicx}                
  \DeclareGraphicsExtensions{.pdf,.png,.jpg,.jpeg} 
\else
  \usepackage{graphicx}                
  \DeclareGraphicsExtensions{.eps}     
\fi%

\graphicspath{{figures/}{pictures/}{images/}{./}} 

\usepackage{microtype}                 
\PassOptionsToPackage{warn}{textcomp}  
\usepackage{textcomp}                  
\usepackage{mathptmx}                  
\usepackage{times}                     
\usepackage{cite}                      
\usepackage{tabu}                      
\usepackage{booktabs}                  
\usepackage{url}
\usepackage{color}
\usepackage{graphics,url,subfigure}
\usepackage{multirow}
\usepackage{float}
\usepackage{amsfonts,amssymb,euscript,mathrsfs,bbm,theorem}
\usepackage{amsmath,scalefnt}
\usepackage{algorithm}
\usepackage{algorithmic}
\usepackage{verbatim}
\usepackage{graphicx,dblfloatfix}
\usepackage[labelsep=period]{caption}
\usepackage{theorem}
\DeclareGraphicsExtensions{.ps}

\definecolor{mycolor}{rgb}{0.00, 0.0, 0.0}
\definecolor{mycolor2}{rgb}{0, 0.5, 0}
\definecolor{mycolor3}{rgb}{0, 0, 0.5}
\definecolor{mycolor4}{rgb}{1, 0, 0}

\newcommand{\ta}[1]{{\color{mycolor}{#1}}}

\newcommand{\delete}[1]{\iffalse{#1}\fi}

\newcommand{\etal}{\textit{et\,al.}}

\newcommand{\FG}[1]{Fig.~\ref{#1}}

\newcommand{\shah}{{\textstyle \amalg{\kern-4.pt\amalg}}}

\newcommand {\mm}[1] {\ifmmode{#1}\else{\mbox{\(#1\)}}\fi}

\newcommand{\PM}{{probabilistic map}}

\onlineid{0}

\vgtccategory{Research}

\vgtcinsertpkg

\title{Statistical Rendering for Visualization of Red Sea Eddy Simulation Data}

\author{Tushar Athawale, Alireza Entezari,~\textit{Senior Member,~IEEE}, \\Bei Wang, and Chris R. Johnson,~\textit{Fellow,~IEEE}}
\authorfooter{
\item
 T. Athawale, L. Yan, B. Wang, and C. R. Johnson are with Scientific Computing \& Imaging (SCI) Institute, University of Utah. E-mail:  \{tushar.athawale, linyan, beiwang, crj\}@sci.utah.edu,
\item
A. Entezari is with the Department of CISE at the University of Florida, Gainesville, FL, 32611. E-mail: entezari@cise.ufl.edu
}

\abstract{Analyzing the effects of ocean eddies is important in oceanology for
gaining insights into transport of energy and biogeochemical particles. We present an application of statistical visualization algorithms for the analysis of the Red Sea eddy simulation ensemble.
Specifically, we demonstrate the applications of \emph{statistical volume rendering} and \emph{statistical
Morse complex summary maps} to a velocity magnitude field for studying the eddy positions in the flow dataset. In statistical volume rendering, we model per-voxel data uncertainty
using noise models, such as parametric and nonparametric, and study the propagation of uncertainty 
into the volume rendering pipeline. In the statistical Morse complex summary maps, we derive histograms charactering uncertainty of gradient flow destinations to understand Morse complex topological variations across the ensemble. We demonstrate the utility of our statistical visualizations for an effective analysis of the potential eddy positions and their spatial uncertainty.%
} 

\CCScatlist{
  \CCScatTwelve{Human-centered computing}{Visualization}{Visualization application domains}{Scientific visualization};
  \CCScatTwelve{Mathematics of computing}{Probability and statistics}{Probabilistic algorithms}{}
}

\begin{document}


\firstsection{Introduction}

\maketitle


In this work, we propose visualizations for the analysis of the Red Sea eddy simulation dataset, which is available via the IEEE SciVis Contest 2020 \footnote{\url{https:
//kaust-vislab.github.io/SciVis2020/}}. The randomness in data acquisition is captured via an ensemble of simulations, in which each ensemble member is generated based on the MIT ocean general circulation model (MITgcm) and the Data Research Testbed (DART)~\cite{HoteitHoarGopalakrishnan2013} with varying initial conditions. Each ensemble is  sampled on a domain with a grid of resolution $500\times500\times50$, and ensembles are sampled for $60$ time steps to represent a time-varying 3D flow~\cite{SivareddyToyeZhan2020}.
 
Although the uncertainty inherent in physical systems may be represented through multiple simulations/sensors, the large size of the ensemble simulations, e.g., 1.5 TB for the Red Sea dataset, can become a bottleneck for data transmission and visualization. The recent work by Wang \etal~\cite{WangHazarikaLi2018} reviewed the challenges as well as advances in visualization methods for handling the complexity of large ensembles. They found that the most common approach to alleviate the complexity of large ensembles is through statistical summarizations. Visualizing statistical summaries of ensembles can help users understand commonalities and differences in the features observed across the ensemble members. Statistical summarizations can also help alleviate the storage burden through reduced representation of data.

We employ two statistical summarization techniques for the visualization of the Red Sea eddy simulation dataset, namely statistical volume rendering~\cite{SakhaeeEntezari2017} and statistical Morse complex summary maps~\cite{AthawaleJohnsonWang2019}. We now demonstrate the results of the application of statistical rendering techniques to the contest dataset.

\section{Statistical 3D Volume Rendering}

We employ statistical volume rendering frameworks~\cite{SakhaeeEntezari2017, LiuLevineBremer2012, AthawaleJohnsonEntezari2021} for 3D visualizations of the Red Sea eddy simulation ensemble. In a statistical volume rendering framework, per-voxel uncertainty is characterized using a probability distribution, which is estimated from the ensemble members. The probability distributions are then propagated through the direct volume rendering pipeline to derive likely (expected) visualizations for the ensemble. 

We derive the expected visualizations for four noise models, namely, uniform~\cite{SakhaeeEntezari2017}, Gaussian, Gaussian mixtures~\cite{LiuLevineBremer2012}, and nonparametric models~\cite{AthawaleJohnsonEntezari2021}, as shown in \FG{fig:kaustStatisticalRenderingNoiseModels}. The visualizations are derived for the velocity magnitude ensemble with $20$ members over the domain 40$^{\circ}$E-50$^{\circ}$E and 10$^{\circ}$N-20$^{\circ}$N for the time step $t=40$. \FG{fig:kaustStatisticalRenderingNoiseModels}a visualizes a single ensemble member using the arrow glyphs color mapped by velocity magnitude. The high-velocity magnitude is generally observed near the vortex rim. The transfer function shown in \FG{fig:kaustStatisticalRenderingNoiseModels}c maps the regions with relatively high-, moderate-, and low-velocity magnitudes to red, blue, and yellow, respectively. The same transfer function is used for all statistical renderings in \FG{fig:kaustStatisticalRenderingNoiseModels}.  

\begin{figure*} [!ht] 
\centering
\includegraphics[width = 6in]{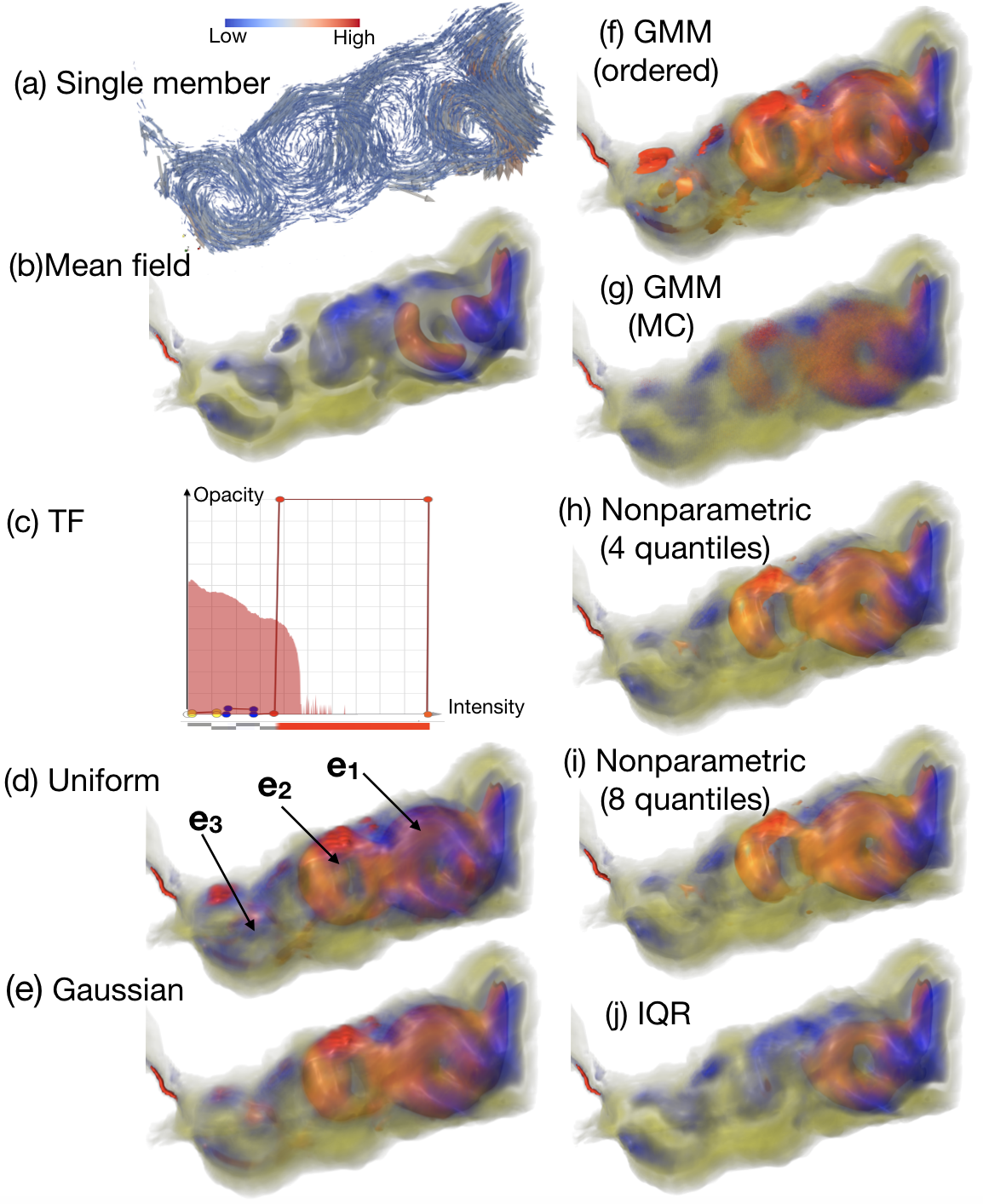}
\caption{\ta{Statistical volume rendering of the velocity magnitude ensemble: (a) arrow glyph visualization for a velocity vector field of a single ensemble member colored by magnitude, (b) mean statistics, (d-e) parametric noise models, (f) Gaussian mixture models with ordered Gaussian means, (g) Gaussian mixture models with Monte Carlo sampling, (h-i) nonparametric density models with quantile representation, (j) uniform noise model for the interquartile range (central 50\% population). The red, blue, and yellow in the transfer function (c) indicate relatively high-, moderate-, and low-velocity magnitudes. $e_1$, $e_2$, and $e_3$ denote the potential eddy positions observed across different noise models.}}
	\label{fig:kaustStatisticalRenderingNoiseModels}
\end{figure*}

The expected visualizations derived using the uniform, Gaussian, Gaussian mixture, and nonparametric statistical models (Fig.~\ref{fig:kaustStatisticalRenderingNoiseModels} (d-j)) appear significantly different from the mean-field visualization (Fig.~\ref{fig:kaustStatisticalRenderingNoiseModels}b). The mean statistics exhibit high sensitivity to the outlier members, thus, they lack reliable reconstructions of the expected vortical features for the ensemble. In contrast, the distribution-based models display reconstructions with relatively high resilience to the outlier members and indicate the presence of eddies in the regions indicated by $e_1$, $e_2$, and $e_3$ (see Fig.~\ref{fig:kaustStatisticalRenderingNoiseModels}d). As can be inferred from the statistical renderings, the eddy denoted by $e_1$ can be observed across all distribution models, thus indicating a high likelihood of its presence/position. The eddy indicated by $e_2$ is clearly seen in the uniform, Gaussian, and Gaussian mixture (ordered) models, but not in the remaining noise models. The eddy denoted by $e_3$ exhibits a high level of uncertainty regarding its presence/position, as no noise model shows a clear vortical structure in the same region.      

Note that the statistical summarizations using the uniform and Gaussian noise models consume only twice the amount of memory needed for the mean-field statistical approach since they store mean and width/variance per voxel. We use four Gaussians for uncertainty modeling with Gaussian mixtures (see~\cite{LiuLevineBremer2012}), which means, they consume $12$ times the amount of memory needed for the mean field (mean, variance, and weight per Gaussian). Quantile interpolation consumes memory proportional to the number of quantiles (see~\cite{AthawaleJohnsonEntezari2021} for more details). The reduced data representation allows for statistical volume rendering at interactive frame rates. (Refer to the supplementary video for the data interaction demo.)

\FG{fig:quartileView}a visualizes a box-plot-like view for the velocity magnitude ensemble. Specifically, we derive the lower quartile (lower 25\%), middle quartile (central 50\%), and upper quartile  (upper 25\%) at each voxel of the dataset and visualize each quartile with the uniform statistics. The quartile view~\cite{AthawaleJohnsonEntezari2021} gives us insight into variations in features across the three populations. \FG{fig:quartileView}b analyzes the effects of sample size on nonparametric statistical renderings. The dotted boxes in \FG{fig:quartileView}b illustrate the features with relatively high sensitivity to underlying data. \FG{fig:kaustStatisticalRenderingTimeEvolution} depicts how visualizations evolve for the time steps $t = 36, \cdots, 40$ for the mean and parametric statistics.

\section{Statistical 2D Morse Complex Summary Maps}

Morse and Morse-Smale complexes are topological descriptors that provide abstract representations of the gradient flow behavior of scalar fields~\cite{EdelsbrunnerHarerZomorodian2001}. We study the variability of Morse complexes for the Red Sea ensemble members using the probabilistic maps~\cite{AthawaleJohnsonWang2019} to extract the expected vortex structures as well as to gain insight into the positional variability of expected vortex structures. For our analysis, we use an ensemble of $10$ members, in which each member corresponds to a 2D slice perpendicular to the z-axis ($z=1$) for time step $40$. We again analyze the eddies over the domain 40$^{\circ}$E-50$^{\circ}$E and 10$^{\circ}$N-20$^{\circ}$N. Each ensemble member represents a velocity vector field, and Morse complexes are computed from the negation of velocity magnitudes of each ensemble member to focus on local minima of the vector fields. The probabilistic map computation comprises three steps: persistence simplification for each member, local maxima association across simplified members via labeling, and Morse complex visualization.

\begin{figure*}[!ht]
\centering 
\includegraphics[width=0.61\textwidth]{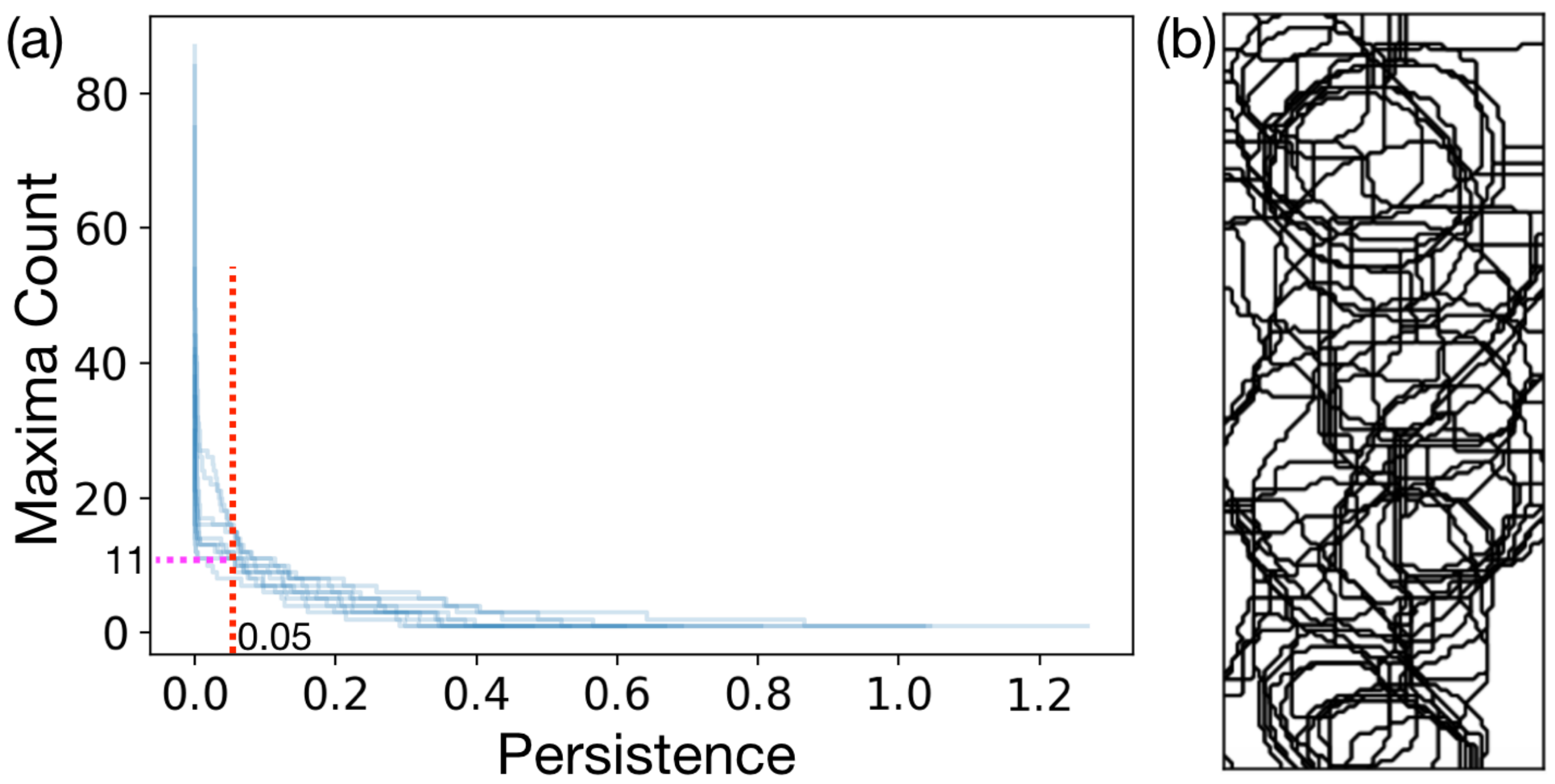}
\caption{Persistence simplification. (a) Persistence graphs. (b) Spaghetti plots of the simplified Morse complexes.}
\label{fig:red-sea-persistence}
\end{figure*}

\paragraph{Persistence simplification.} 
Persistent homology is a tool in topological data analysis for quantifying
the significance of topological features. It is widely used for data de-noising through persistence simplification~\cite{EdelsbrunnerLetscherZomorodian2002}. We employ persistence simplification to obtain a common label set across all ensemble members, guided by persistence graphs and spaghetti plots in Fig.~\ref{fig:red-sea-persistence}. 
In particular, at the selected simplification scale (dotted red line) in Fig.~\ref{fig:red-sea-persistence}a, $5$ of $10$ ($50\%$) members agree on the number of maxima ($11$) after simplification. 

\begin{figure*}[!ht]
  \centering
  \includegraphics[width=0.61\textwidth]{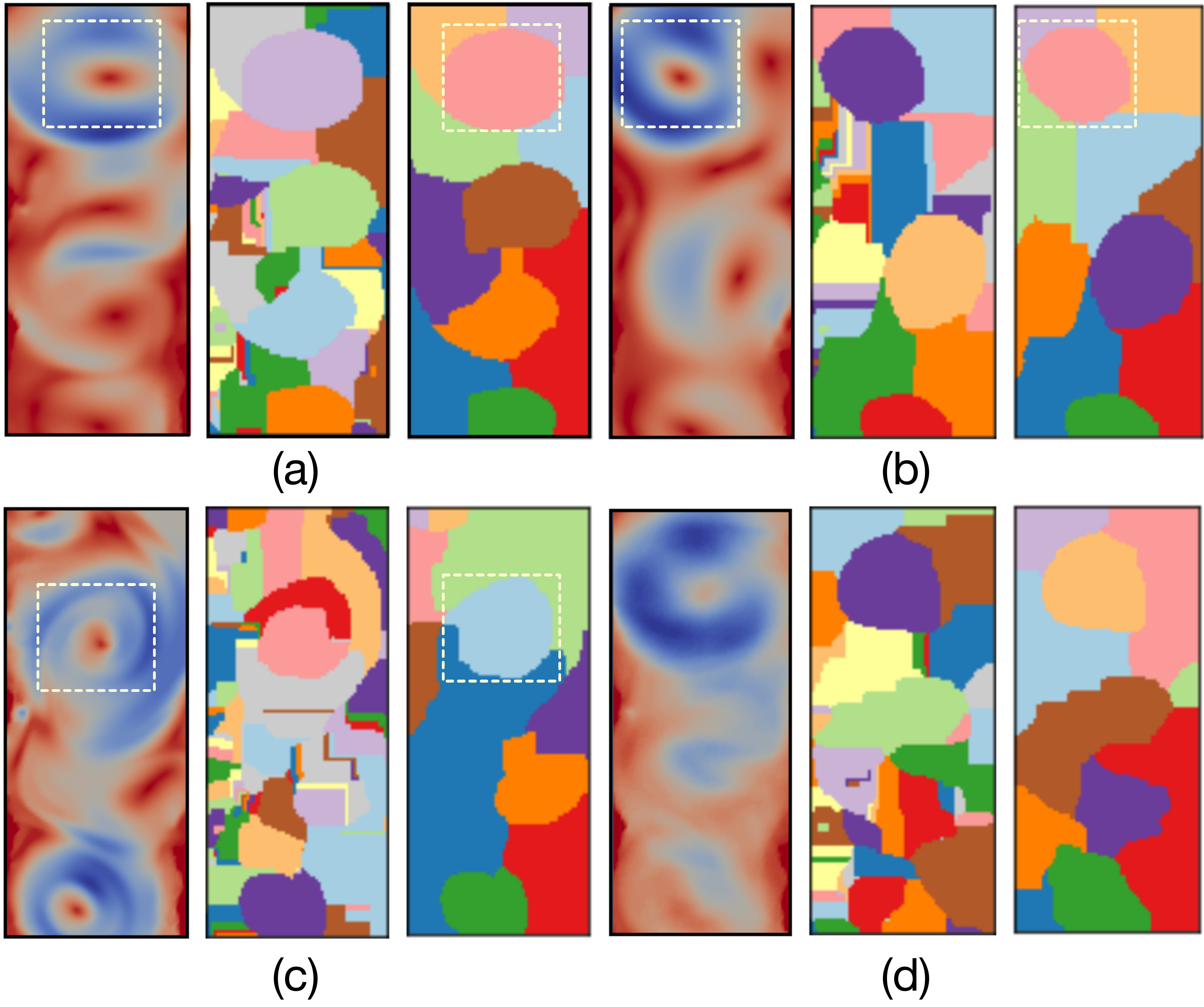}
    \caption{(a-c) Three ensemble members together with (d) the mean field. Each subfigure visualizes, from left to right, the negated velocity magnitude field (red means low and blue means high velocity magnitude), and its corresponding Morse complexes before and after persistence simplification. The dotted white boxes mark the vortex features of each member.}
    \label{fig:red-sea-mean}
\end{figure*}

We illustrate three ensemble members in Figs.~\ref{fig:red-sea-mean}a-c, respectively. 
For each ensemble member, its corresponding simplified Morse complex contains 2-cells that highlight vortical features of ocean eddies (white boxes). The mean field Morse complex in Fig.~\ref{fig:red-sea-mean}d, however, does not give any insight into the structural uncertainty, that is, the variabilities of these features across the ensemble. The spaghetti plots of the simplified Morse complexes in Fig.~\ref{fig:red-sea-persistence}b do not display the topological consistency of 1-cells, thereby indicating the high variability of simulations. For the simulations with high variability, we benefit from the k-means and Morse mapping labeling strategies for deriving associations among local maxima of ensemble members, as demonstrated below. 

\begin{figure*}[!ht]
  \centering
    \includegraphics[width=0.61\textwidth]{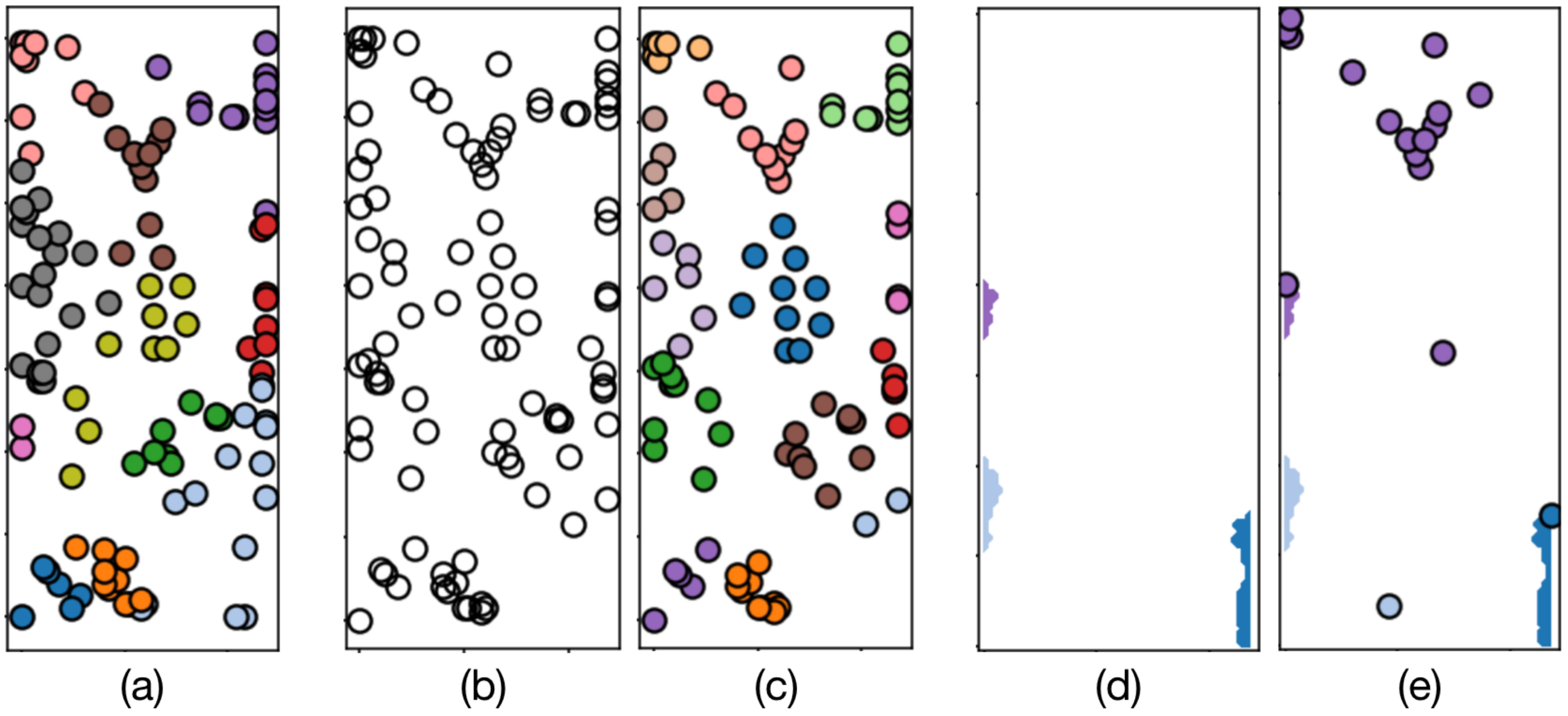}
    \caption{(a) Labeling with Morse mapping. (b-c) Labeling with k-means clustering. (d) Mandatory maxima are shown as colored regions. (e) Labeling with nearest mandatory maxima. }
    \label{fig:red-sea-labeling}
\end{figure*}

\paragraph{Labeling.}
In Fig.~\ref{fig:red-sea-labeling}, we compare the three labeling strategies proposed in~\cite{AthawaleJohnsonWang2019}.  
As illustrated in Fig.~\ref{fig:red-sea-labeling}d, the number of mandatory maxima~\cite{DavidJosephJulien2014} is small ($3$) since ensemble members have large variations. 
Simplifying each ensemble member to have $3$ maxima will miss most of the features of interest (Fig.~\ref{fig:red-sea-labeling}e).  
The Morse mapping (Fig.~\ref{fig:red-sea-labeling}a) and the k-means clustering (Fig.~\ref{fig:red-sea-labeling}b-c) strategies, on the other hand, provide reasonable results. In the k-mean clustering, we set $k=11$ since we simplified each ensemble member to contain $11$ maxima based on the analysis of persistence graphs. The Morse mapping is more flexible than the k-means without requiring the same number of maxima across the ensemble. 

\begin{figure*}[!ht]
  \centering 
    \includegraphics[width=0.72\textwidth]{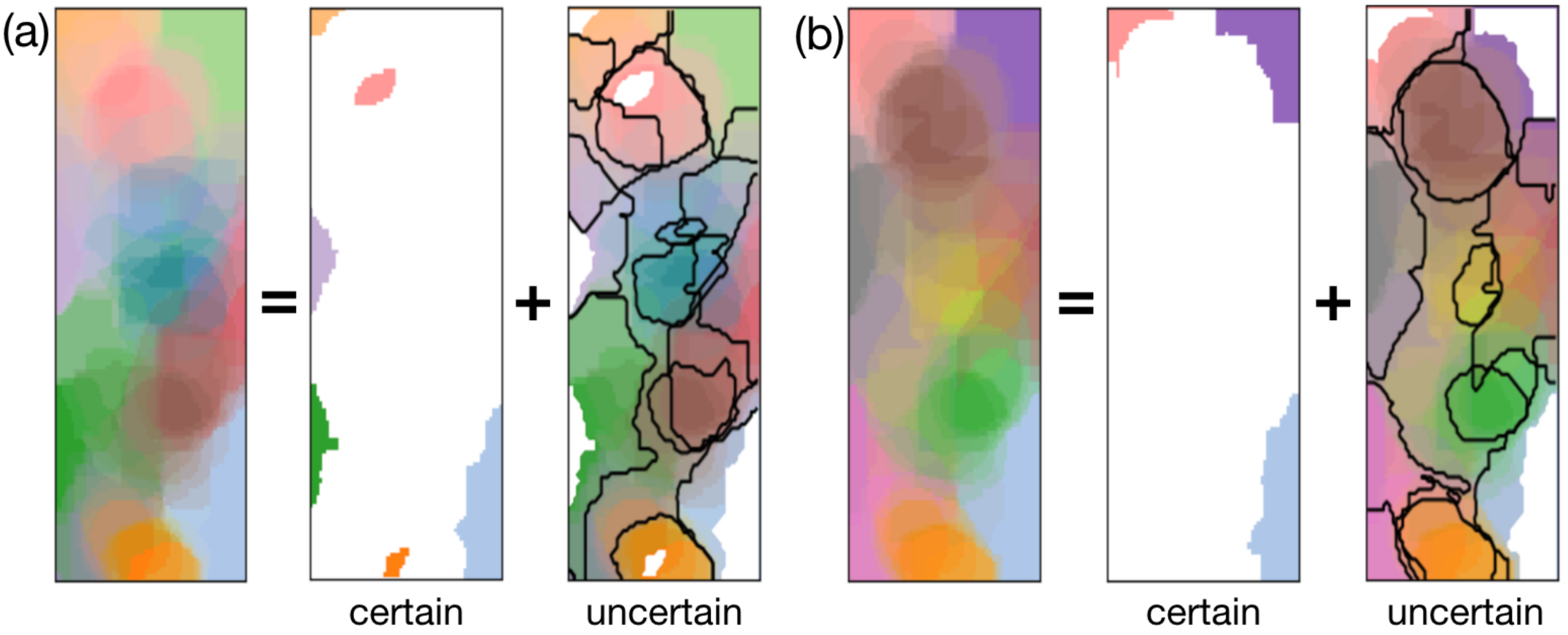}
    \caption{The probabilistic map is visualized based on (a) k-means clustering and (b) Morse mapping strategies.}
    \label{fig:red-sea-probability}
\end{figure*}

\begin{figure*}[!ht]  
  \centering
  \includegraphics[width=0.72\textwidth]{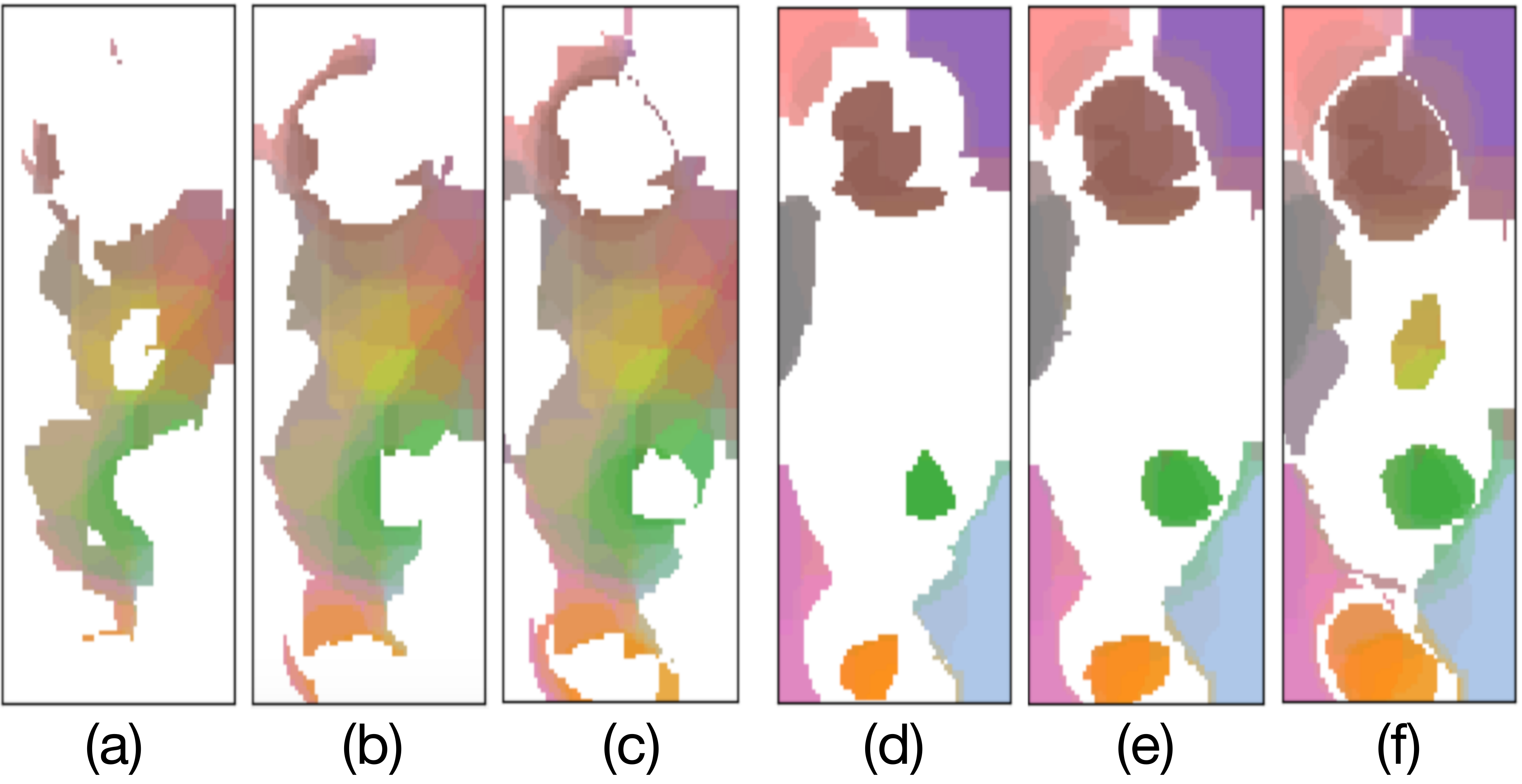}
    \caption{(a-c) Entropy-based exploration of uncertain regions representing entropy greater than or equal to $1.5$ in (a), $1.25$ in (b), and $1$ in (c), respectively; (d-f) visualizations of the regions that agree in their gradient destinations for at least 80\% members in (d), 70\% members in (e), and 60\% members in (f), respectively.}
    \label{fig:red-sea-entropy-agreement-exploration}
\end{figure*}

\begin{figure*}[!ht]  
  \centering
  \includegraphics[width=0.72\textwidth]{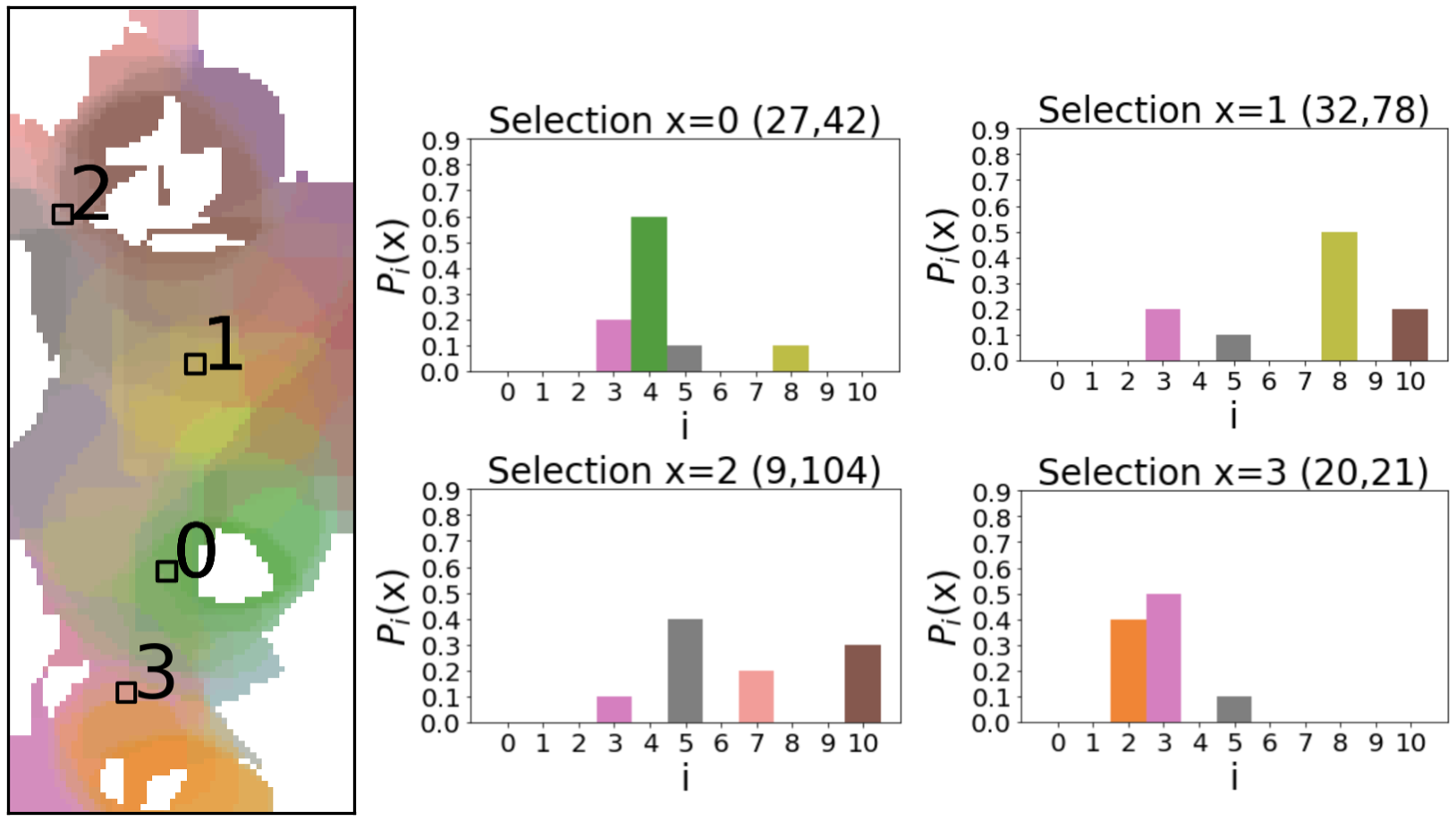}
    \caption{The probabilistic map for the positions with entropy greater than or equal to $0.8$. The query selections $0-3$ have the highest probability of flowing to the maxima with green, yellow, gray, and pink labels, respectively.}
    \label{fig:red-sea-interactive-queries}
\end{figure*}

\paragraph{Probabilistic map.} 
We visualize the {\PM} using color blending~\cite{AthawaleJohnsonWang2019} for both k-means clustering and Morse mapping labeling strategies.
Both visualizations in Fig.~\ref{fig:red-sea-probability} highlight the positional uncertainty of 2-cell boundaries invisible to the mean field of Fig.~\ref{fig:red-sea-mean}d. 
However, the expected 2-cell boundaries (black contours) using Morse mapping appear to be more spatially stable than those obtained via k-means clustering. The expected 2-cell boundaries extract the expected eddy positions for the ensemble dataset. Figs.~\ref{fig:red-sea-entropy-agreement-exploration}a-c visualize our entropy-based exploration of the {\PM} for lower entropy thresholds of $1.5$, $1.25$, and $1$, respectively. Figs.~\ref{fig:red-sea-entropy-agreement-exploration}d-f carve out regions in the domain, where the ensemble agrees in their gradient destinations for at least 80\%, 70\%, and 60\% members, respectively. Thus, the shared features denoting the eddy structures across the ensemble are discoverable in Figs.~\ref{fig:red-sea-entropy-agreement-exploration}d-f. In Fig.~\ref{fig:red-sea-interactive-queries}, the probabilistic map is again visualized for the lower entropy threshold of $0.8$. The gradient flows originating at the query selections $0-3$ in Fig.~\ref{fig:red-sea-interactive-queries} have the highest probability of terminating at the local maxima with green, yellow, gray, and pink labels, respectively.

\section{Implementation}
In the case of statistical volume visualizations, the renderings are performed on a machine with Nvidia GPU Quadro
P6000, with 24 GB memory. We integrated the fragment shaders for our statistical frameworks into the Voreen volume rendering engine (\url{http://voreen.uni-muenster.de}) for DVR of ensemble data. In the case of statistical Morse complex summary maps, we extend the Python code for topological data analysis available at \url{https://pypi.org/project/topopy/}. We provide the demo of our techniques in action in a supplementary video.

\section{Conclusion}
We demonstrate the effectiveness of statistical visualization techniques for aggregate analysis of the Red Sea eddy simulation dataset. Specifically, we illustrate applications of statistical volume rendering~\cite{SakhaeeEntezari2017, LiuLevineBremer2012, AthawaleJohnsonEntezari2021} and statistical Morse complex summary maps~\cite{AthawaleJohnsonWang2019} to extract the likely (expected) eddy positions as well as their variability. The distribution-based data representation in the case of statistical volume rendering allows for the exploration of the large-scale Red Sea eddy simulation ensemble in 3D at interactive frame rates. Additionally, the distribution-based statistics show increased robustness to noise compared to the mean statistics and allow for an uncertainty integration with visualizations using both statistical rendering techniques.  

\acknowledgments{
This work was supported in part by the NIH grants P41 GM103545-18 and R24
GM136986; the DOE grant DE-FE0031880; the Intel
Graphics and Visualization Institutes of XeLLENCE; and the NSF grants IIS-1617101, IIS-1910733, DBI-1661375,
and IIS-1513616.}

\bibliography{template}

\newpage

\begin{figure*} [!ht] 
\centering
\includegraphics[width = 6in]{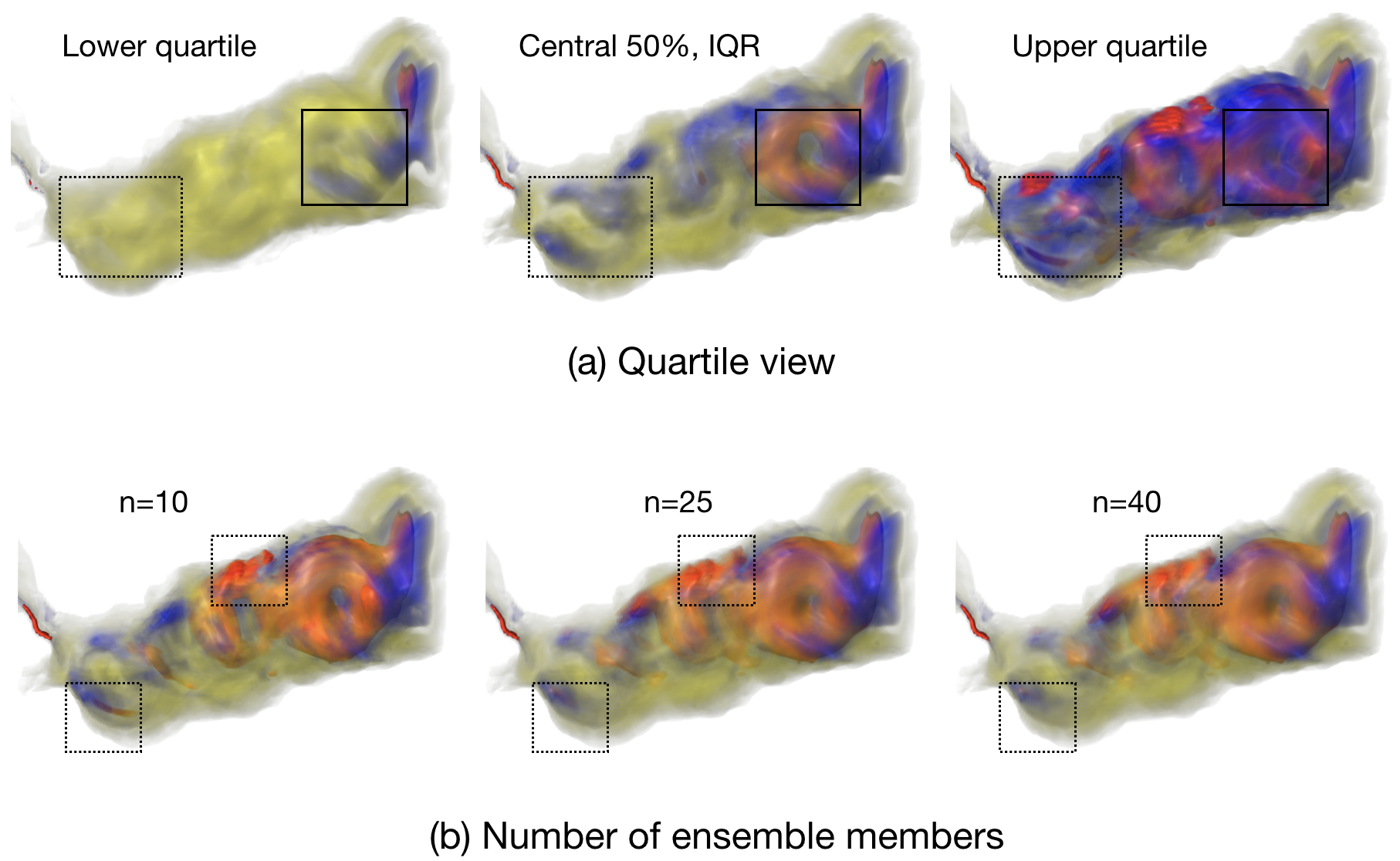}
\caption{(a) The quartile view for uncertainty analysis of an eddy presence in
the Red Sea dataset. The solid boxes enclose the positions that indicate
the high likelihood of an eddy presence, whereas the dotted boxes mark
positions with substantial uncertainty in an eddy presence. (b) Effect of sample size (n) on visualizations. The dotted boxes illustrate positions with variability in reconstruction.}
	\label{fig:quartileView}
\end{figure*}

\begin{figure*} [!ht] 
\centering
\includegraphics[width = 6in]{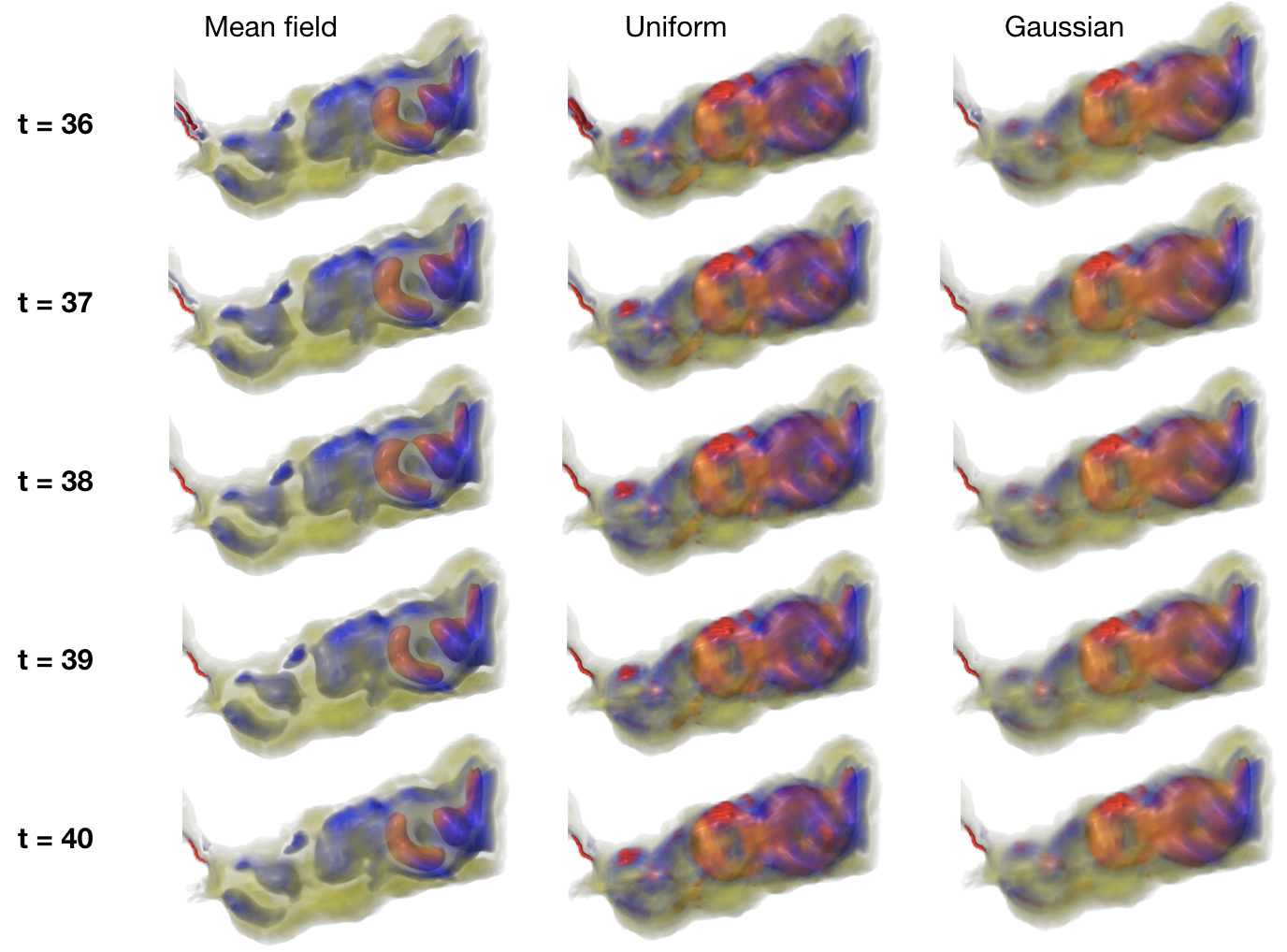}
\caption{Visualizations of the uncertain velocity magnitude field for a series of time steps. The mean-field visualization exhibits relatively more fluctuations in reconstructions compared to the uniform and Gaussian noise models.}
	\label{fig:kaustStatisticalRenderingTimeEvolution}
\end{figure*} 

\newpage

\bibliographystyle{abbrv-doi}

\end{document}